\documentclass[journal]{IEEEtran}

\usepackage[cmex10]{amsmath}
\usepackage{amssymb}
\usepackage{textcomp}
\usepackage[pdftex]{graphicx}
\usepackage{cite}
\usepackage{hyperref}
\usepackage{xcolor}

\usepackage[inline]{enumitem}

\usepackage{mathtools}  

\begin{document}
\title{Hardware Implementation of Fano Decoder for Polarization-adjusted Convolutional (PAC) Codes}

\author{Amir Mozammel\textsuperscript{\href{https://orcid.org/0000-0003-3474-9530}{\includegraphics[scale=0.06]{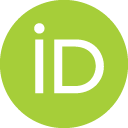}}},~\IEEEmembership{Student Member,~IEEE}
\thanks{The author is with the Department of Electrical-Electronics Engineering, Bilkent University, Ankara TR-06800, Turkey (e-mail: a.mozammel@ee.bilkent.edu.tr).}
}

\maketitle

\begin{abstract}
This brief proposes a hardware implementation architecture for Fano decoding of polarization-adjusted convolutional (PAC) codes.
This architecture uses a novel branch metric unit specific to PAC codes.
The proposed decoder is tested on FPGA, and its performance is evaluated on ASIC using TSMC 28 nm 0.72 V library.
The decoder can be clocked at 500 MHz and reach an average information throughput of 38 Mb/s at 3.5 dB signal-to-noise ratio for a block length of 128 and a code rate of 1/2.
\end{abstract}


\IEEEpeerreviewmaketitle

\section{Introduction}\label{sec_introduction}
\IEEEPARstart{T}{his} brief presents a hardware implementation study of Fano decoding for polarization-adjusted convolutional (PAC) codes, which are a new class of error-correcting codes introduced in \cite{arikan2019sequential}.
PAC codes combine ideas from polar coding \cite{arikan_polar} and convolutional coding and have been shown to perform near the dispersion approximation \cite{polyanskiy} in certain cases \cite{arikan2019sequential}.
Fano decoding \cite{fano1963heuristic} is a depth-first tree search algorithm which was originally developed for convolutional codes.
PAC codes can be decoded using any tree search algorithm such as depth-first, breadth-first, and beam (constrained breadth-first) search algorithms.

Depth-first search decoders for convolutional codes are in general known as sequential decoding \cite{wozencraft} algorithms.
Two well-known sequential decoders are Fano and stack algorithms \cite{zigangirov, jelinek}.
Compared to stack algorithm, Fano decoder requires a smaller memory size and is more suitable for hardware implementations.
For this reason, in this brief, we focus on the Fano version of sequential decoding.
Various architectures for hardware implementation of sequential decoding of convolutional codes have been reported in the literature \cite{benaissa2007reconfigurable,forney1971high,jacobs1967sequential,layland1971flexible}, but to the best of the author's knowledge, the suitability of sequential decoding for PAC codes has never been studied from a hardware implementation perspective.
Motivated by this, we implement a Fano decoder for PAC codes by introducing a new hardware-friendly variant of Fano algorithm for PAC codes and designing a novel branch metric unit capable of calculating the current and previous branch metrics without requiring any storage element or comparator.

Despite its near-optimal performance, PAC codes under sequential decoding exhibit variable time complexity, resulting in variable decoding latency.
Although the depth-first search algorithms have variable search complexity, their average search complexity is low at a high signal-to-noise ratio (SNR) regime.
On the other hand, breadth-first search algorithms have fixed but higher search complexity.
List decoding \cite{vardi2021list} of PAC codes is an example of beam search decoder.
However, the list decoder requires a large list size to achieve the error-correction performance of the PAC sequential decoder.

Throughout this brief, we denote vectors by boldface letters.
For any set $\mathcal{A} \subseteq \{0,1,...,N-1\}$, we denote its complement by $\mathcal{A}^c = \{i: i \notin A \}$.
For any vector $\mathbf{y}$ and set $\mathcal{A}$, $\mathbf{y}_\mathcal{A}$ denotes the sub-vector $(y_i : i \in \mathcal{A})$.
For any vector $\mathbf{y}$, $\mathbf{y}^j = (y_0,y_2,...,y_j)$.
We define a sign function $s(l)$ such that $s(l) = 0$ if $l \geq 0$ and $s(l) = 1$, otherwise.

The rest of this brief is organized as follows.
Section~\ref{sec_pac} gives a brief discussion of PAC codes.
Section~\ref{sec_fano_algo} introduces a new variant of Fano decoder for PAC codes.
In Section~\ref{sec_architecture} we introduce a hardware architecture for Fano decoding of PAC codes.
Implementation results of the proposed PAC Fano decoder are presented in Section~\ref{sec_results}.
Section~\ref{sec_conclusion} concludes this brief.

\begin{figure}[t]
\centering
\includegraphics[width=0.9\columnwidth]{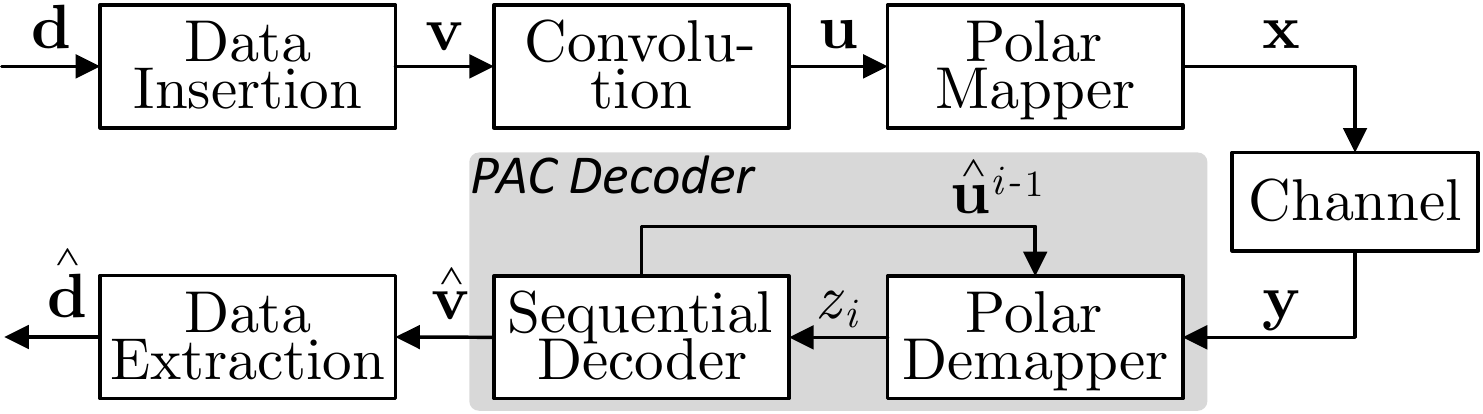}
\caption{PAC coding scheme.}
\label{fig_pac_scheme}
\end{figure}
\section{PAC Codes}\label{sec_pac}
Fig.~\ref{fig_pac_scheme} shows a block diagram of PAC coding scheme.
The data insertion block receives a source word $\mathbf{d}$ of length $K$ and inserts it into a data carrier word $\mathbf{v}$ of length $N$ in accordance with a data index set $\mathcal{A}$ such that $\mathbf{v}_\mathcal{A} = \mathbf{d}$ and $\mathbf{v}_{\mathcal{A}^c} = \mathbf{0}$.
The bits fixed to zero are called frozen, whereas all the other bits are called non-frozen.
The data carrier word $\mathbf{v}$ goes through a convolution block with generator matrix $\mathbf{G}$ which is a Toeplitz matrix whose first row is $\mathbf{c}$ (the generator polynomial).
The resulting word $\mathbf{u}$ goes through a polar mapper and the overall encoding process of PAC codes can be expressed by $\mathbf{x}=\mathbf{v}\mathbf{G}\mathbf{F}^{\otimes n}$, where $\mathbf{F}^{\otimes n}$ is the generator matrix of polar codes with $\mathbf{F}^{\otimes n}$ being the $n$-th Kronecker product of the kernel matrix $\mathbf{F} = \begin{bsmallmatrix} 1 & 0\\ 1 & 1 \end{bsmallmatrix}$.  

At the receiver side, the PAC decoder receives the channel output $\mathbf{y}$ and generates an estimate $\hat{\mathbf{v}}$ of $\mathbf{v}$.
Then, a data extractor extracts an estimate $\hat{\mathbf{d}}$ of $\mathbf{d}$ from $\hat{\mathbf{v}}$ using $\hat{\mathbf{d}} = \hat{\mathbf{v}}_\mathcal{A}$.
The performance of the system is measured by the probability of frame error $P_e = P(\hat{\mathbf{d}} \neq \mathbf{d})$.
A PAC sequential decoder consists of two blocks: polar demapper and sequential decoder.
The polar demapper receives the channel output $\mathbf{y}$ and calculates a log-likelihood ratio (LLR) vector $\boldsymbol{l}$.
Then, based on the prior bit-estimates $\hat{\mathbf{u}}^{i-1}$ received from sequential decoder, it generates the demapped LLR value $z_i \triangleq \ln{\frac{P(\mathbf{y},\hat{\mathbf{u}}^{i-1}|\hat{u}_i=0)}{P(\mathbf{y},\hat{\mathbf{u}}^{i-1}|\hat{u}_i=1)}}$ of $i$th bit.
Polar demapper operates similar to successive cancellation (SC) decoder of polar codes with a difference that the polar demapper does not generate any bit-estimate output $\hat{u}_i$.
Instead, it receives the prior bit-estimates $\hat{\mathbf{u}}^{i-1}$ from the sequential decoder and passes back the soft value of $z_i$.
The sequential decoder uses $z_i$ to calculate a path metric which helps the decoder to generate an estimate $\hat{\mathbf{v}}$ of $\mathbf{v}$.

\section{Fano Algorithm for PAC Codes}\label{sec_fano_algo}
Fano algorithm uses a path metric and a metric threshold $T$ to identify the correct path in the code tree.
The threshold can only take integer multiples of threshold spacing $\Delta$.
If the path metric grows along a given path, the algorithm considers it as a correct path and continues to search further along it.
But if the metric drops significantly, the algorithm moves back (backtracks) and searches other paths.
Upon a backward move, if the currently reached node is frozen or all of its children are examined, the decoder moves back one more.
An extensive study of the Fano decoder may be found in \cite{gallager1968information}.

\begin{figure}[t]
\centering
\includegraphics[width=0.8\columnwidth]{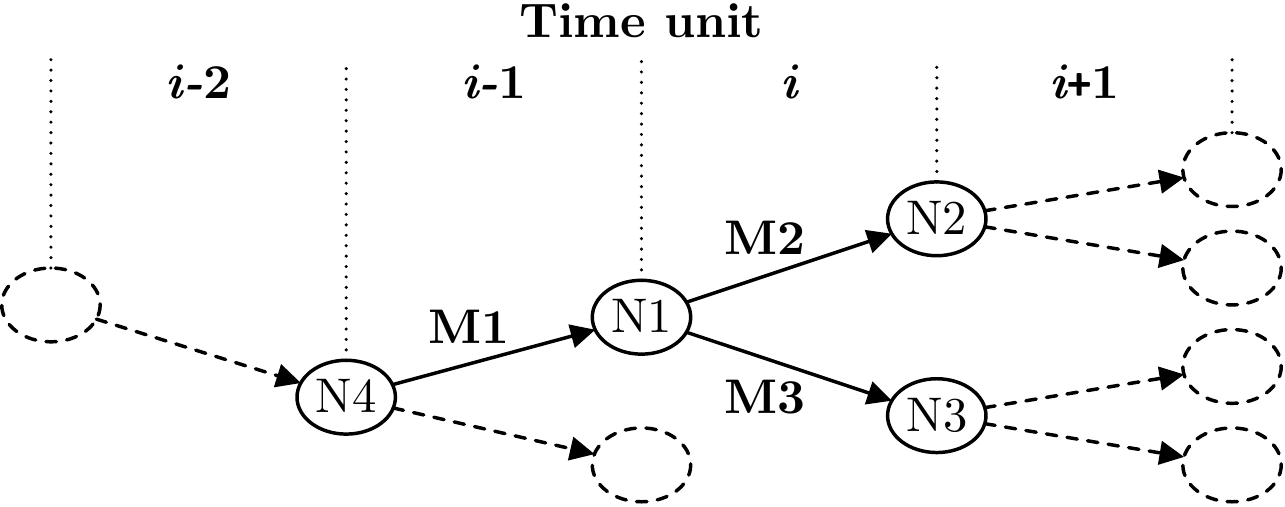}
\caption{Fano decoding tree, reproduced from \cite{benaissa2007reconfigurable}.}
\label{fig_fano_tree}
\end{figure}

Fig.~\ref{fig_fano_tree} shows a local node diagram of the Fano decoding tree.
Assume that N1 is the current node, N2 is the most likely node (with larger metric), N3 is the least likely node (with smaller metric), and N4 is the previous node.
M2 and M3 denote the metrics of branches from current node to N2 and N3, respectively, and M1 is the metric of the branch from previous node to the current node.
To avoid branch metric overflow, we use relative branch metric computation instead of absolute branch metric value \cite{benaissa2007reconfigurable}.
Adopting this method lets the current node's branch metric be zero and all other branch metrics relative to the current node's metric.

We modify the conventional Fano algorithm to make it suitable for decoding PAC codes and state it as the following set of rules.
We define a variable $\Psi$ such that $\Psi = 1$ when the Fano decoder backtracks to a frozen node or to a node whose both children are examined; otherwise, $\Psi = 0$.
N23 is a node that corresponds to N2 when the most likely node is being examined or N3 when the least likely node is being examined.
Similarly, M23 corresponds to M2 when N2 is being examined or M3 when N3 is being examined.
A node is considered as a new node if it is being visited for the first time.

\noindent$\bullet$ \textit{Rule 0}\\
\textbf{Conditions:} $\Psi = 0$, N23 is new node, $\text{M23} \geq T$.\\
\textbf{Actions:} Move to N23, update $T$ to $T + \Delta - \text{M23}$, examine the most likely node leading from N23 at the next step.\\
\noindent$\bullet$ \textit{Rule 1}\\
\textbf{Conditions:} $\Psi = 0$, N23 is old node, $\text{M23} \geq T$.\\
\textbf{Actions:} Move to N23, update $T$ to $T - \text{M23}$, examine the most likely node leading from N23 at the next step.\\
\noindent$\bullet$ \textit{Rule 2}\\
\textbf{Conditions:} $\Psi = 0$, $\text{M23} < T$, N1 is root node; or $\Psi = 0$, $\text{M23} < T$, N1 is not root node, $\text{M1} + T > 0$; or $\Psi = 1$, N1 is root node; or $\Psi = 1$, N1 is not root node, $\text{M1} + T > 0$\\
\textbf{Actions:} Make no move, update $T$ to $T - \Delta$, assign $\Psi = 0$, examine N2 again at the next step\\
\noindent$\bullet$ \textit{Rule 3}\\
\textbf{Conditions:} $\Psi = 0$, $\text{M23} < T$, N1 is not root node, $\text{M1} + T \leq 0$, N4 is not frozen, N1 is the most likely node leading from N4; or $\Psi = 1$, N1 is not root node, N4 is not frozen, N1 is the most likely node leading from N4.\\
\textbf{Actions:} Move to N4, update $T$ to $T + \text{M1}$, assign $\Psi = 0$, examine the lateral node of N1 at the next step.\\
\noindent$\bullet$ \textit{Rule 4}\\
\textbf{Conditions:} $\Psi = 0$, $\text{M23} < T$, N1 is not root node, $\text{M1} + T \leq 0$, N4 is frozen or N1 is the least likely node leading from N4; or $\Psi = 1$, N1 is not root node, N4 is frozen or N1 is the most likely node leading from N4.\\
\textbf{Actions:} Move to N4, update $T$ to $T + \text{M1}$, assign $\Psi = 1$, perform backward check at the next step.\\

\begin{figure}[t]
\centering
\includegraphics[width=0.6\columnwidth]{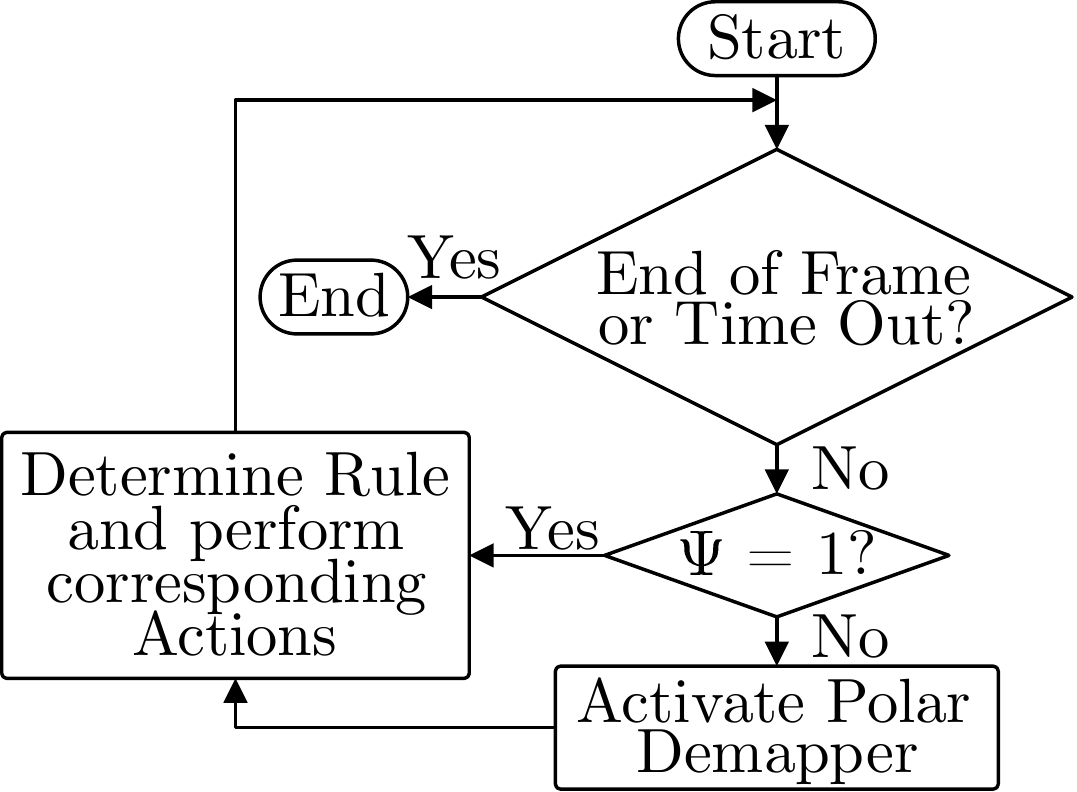}
\caption{Flowchart of Fano rules.}
\label{fig_fano_flowchart}
\end{figure}

Fig.~\ref{fig_fano_flowchart} shows the corresponding flowchart of the Fano decoder.
Note that whenever $\Psi = 1$, the Fano algorithm performs a backward check and does not require any new $z_i$ value.
Thus, by storing the previously generated $z_i$ values, we can avoid activating the polar demapper when $\Psi = 1$ and significantly reduce the decoder's latency (especially at low SNR regime).

\section{Architecture of Fano Decoder for PAC Codes} \label{sec_architecture}
\begin{figure}[t]
\centering
\includegraphics[width=1.0\columnwidth]{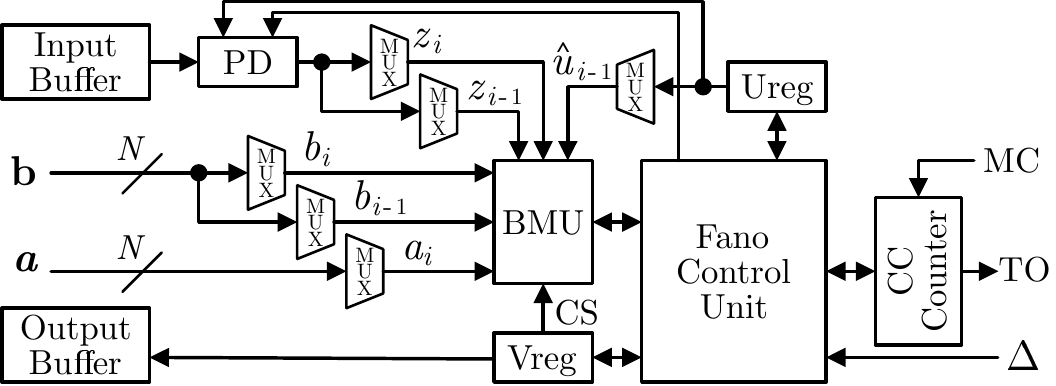}
\caption{PAC Fano decoder.}
\label{fig_decoder}
\end{figure}

Fig.~\ref{fig_decoder} shows the hardware architecture of the proposed PAC Fano decoder.
The input buffer stores the channel output LLR values, and the output buffer stores the final estimate $\hat{\mathbf{v}}$ of $\mathbf{v}$.
The Fano control unit (FCU) implements the flowchart of Fig.~\ref{fig_fano_flowchart}.
The branch metric unit (BMU) is responsible for providing the FCU with the current branch metric M23 and previous branch metric M1.
Vreg is a bidirectional shift register used to store the prior convolution input estimates $\hat{\mathbf{v}}$.
Whenever the Fano decoder moves forward, the current convolution input estimate $\hat{v}_i$ is stored in Vreg.
To allow a maximum backtracking depth of $N$, the size of Vreg is chosen to be $N+h$, where $h$ is the memory size of the convolution.
The first $h$ part of Vreg provides the convolution state (CS) for the BMU.
The Ureg register is used to store the prior convolution output estimates $\hat{\mathbf{u}}$.
When the Fano decoder moves forward, depending on the proceeding branch, 
the corresponding $\hat{u}_i$ is stored in the Ureg.
A clock cycle (CC) counter is used to count the number of clock cycles consumed for decoding a single codeword.
The decoding of a codeword is terminated whenever the value of the CC counter exceeds a predefined maximum cycle (MC).
In this case, a timeout (TO) signal is generated, and a new LLR vector $\boldsymbol{l}$ is loaded to the input buffer.
The input $\boldsymbol{a}$ determines the frozen and non-frozen nodes such that $a_i = 0$ for frozen nodes ($i \in \mathcal{A}^c$) and $a_i = 1$ for non-frozen nodes ($i \in \mathcal{A}$).

To implement the polar demapper (PD), we adopt the FFT-like architecture of \cite{leroux2011hardware} and apply the following modifications to make it operate as a polar demapper:
\begin{enumerate*}
    \item We remove the decision unit and pass the soft values of LLR vector $\mathbf{z}$ to the output;
    \item We remove the bit-estimates update unit and corresponding registers and implement the bit-estimates update network using a combinational circuit that receives $\hat{\mathbf{u}}$ from Ureg and updates the intermediate bit-estimate values;
    \item We modify the bit-reversal architecture of \cite{leroux2011hardware} to output the LLR values in a natural order.
\end{enumerate*}
Despite the similarity in architecture, the timing schedule of PD is different from the one of SC decoder.
Once the PD generates the LLR value of a node $z_i$, it remains idle until another node LLR value is requested.
Note that the next LLR value request may be for the immediate forward node or any other backward node.
Hence, the PD must be able to follow the Fano algorithm whenever it backtracks.
To fulfill this requirement, all the intermediate LLR values are required to be stored and kept until the end of each decoding session.
The FFT-like architecture of \cite{leroux2011hardware} uses distributed registers to stores the intermediate LLR values.
Any SC decoder that is capable of storing the intermediate LLR values can be used as a polar demapper.

The BMU block is a fundamental block which makes the Fano decoding of PAC codes different from Fano decoding of convolutional codes.
The metric function that BMU of PAC Fano decoder uses must be compatible with the polarized channel (created by the polar mapper and demapper) seen by the convolution block of PAC codes.
For PAC codes, the well-known branch metric function of Fano \cite{fano1963heuristic,massey_fan_metric} becomes
\begin{equation}
\gamma_i(\hat{u}_i) = \log_2\frac{P(\mathbf{y},\hat{\mathbf{u}}^{i-1}|\hat{u}_i)}{P(\mathbf{y},\hat{\mathbf{u}}^{i-1})}-b_i,
\label{eqn_pac_branch_metric}
\end{equation}
where $P(\mathbf{y},\hat{\mathbf{u}}^{i-1}|\hat{u}_i)$ and $P(\mathbf{y},\hat{\mathbf{u}}^{i-1})$ are transition and output probabilities of the $i$th bit-channel, respectively, and $b_i$ is a bias term.
For a binary input channel with uniform input distribution, we have
\begin{equation}\label{eqn_totalprob}
    P(\mathbf{y},\hat{\mathbf{u}}^{i-1}) = \frac{1}{2} \left[P(\mathbf{y},\hat{\mathbf{u}}^{i-1}|\hat{u}_i = 0) +P(\mathbf{y},\hat{\mathbf{u}}^{i-1}|\hat{u}_i = 1)  \right].
\end{equation}
By using \eqref{eqn_totalprob} and dividing the numerator and denominator of the fractional part of \eqref{eqn_pac_branch_metric} by $P(\mathbf{y},\hat{\mathbf{u}}^{i-1}|\hat{u}_i)$, after some calculus, we obtain
\begin{equation}
\gamma_i(\hat{u}_i) = 1-\log_2(1+e^{-(1-2\hat{u}_i) z_i})-b_i,
\label{eqn_pac_branch_metric_bi}
\end{equation}
where $z_i$ is the output of PD.
To obtain a hardware-friendly version of \eqref{eqn_pac_branch_metric_bi}, we apply the approximation $\log_2(1+e^{-(1-2\hat{u}_i) z_i}) \approx 0$, if $\hat{u}_i = s(z_i)$, and $\log_2(1+e^{-(1-2\hat{u}_i) z_i}) \approx |z_i|$, otherwise, and obtain
\begin{align}
\gamma_i(\hat{u}_i) =
\begin{cases}
    1-b_i, &\text{if}~\hat{u}_i = s(z_i), \\
    1-|z_i|-b_i, &\text{otherwise}.
\label{eqn_branch_met_simple}
\end{cases}
\end{align}
To simplify \eqref{eqn_branch_met_simple} further, we assume $b_i$ can take only binary values.
As a result, we can tabulate $\gamma_i(0)$ and $\gamma_i(1)$ for all the possible values of $b_i$ and $s(z_i)$ in Table~\ref{tab_gamma}.

\begin{figure}[t]
\centering
\includegraphics[width=.75\columnwidth]{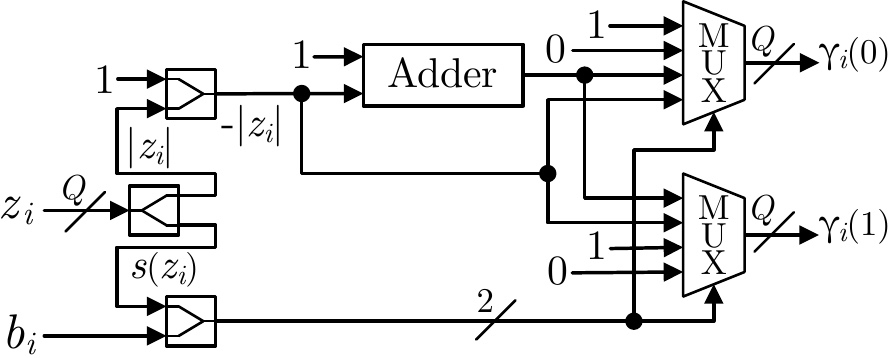}
\caption{Hardware implementation of Table \ref{tab_gamma} (metric calculator).}
\label{fig_metric_calculator} 
\end{figure}

\begin{table}[htpb]
\centering
\caption{$\gamma_i(\hat{u}_i)$ for Different Values of $s(z_i)$ and $b_i$.}
\label{tab_gamma}
\begin{tabular}{cccc}
\hline
$s(z_i)$ & $b_i$ & $\gamma_i(0)$ & $\gamma_i(1)$ \\ \hline
0         & 0     & 1             & $1-|z_i|$     \\ 
0         & 1     & 0             & $-|z_i|$      \\ 
1         & 0     & $1-|z_i|$     & 1             \\ 
1         & 1     & $-|z_i|$      & 0             \\ \hline
\end{tabular}
\end{table}

We can implement this table using two 4-to-1 multiplexers and one adder.
Fig.~\ref{fig_metric_calculator} shows the hardware implementation of Table \ref{tab_gamma} (metric calculator).
The number of quantization bits for LLRs is denoted by $Q$.
The metric calculator receives $z_i$ and $b_i$ and generates the branch metric $\gamma_i(\hat{u}_i)$ for the two possible values of $\hat{u}_i=0$ and $\hat{u}_i=1$.
The constant '0' and '1' inputs to the adder and multiplexers are padded with zeros to have $Q$-bit width (not shown in the figure for clarity).

\begin{figure}[t]
\centering
\includegraphics[width=0.9\columnwidth]{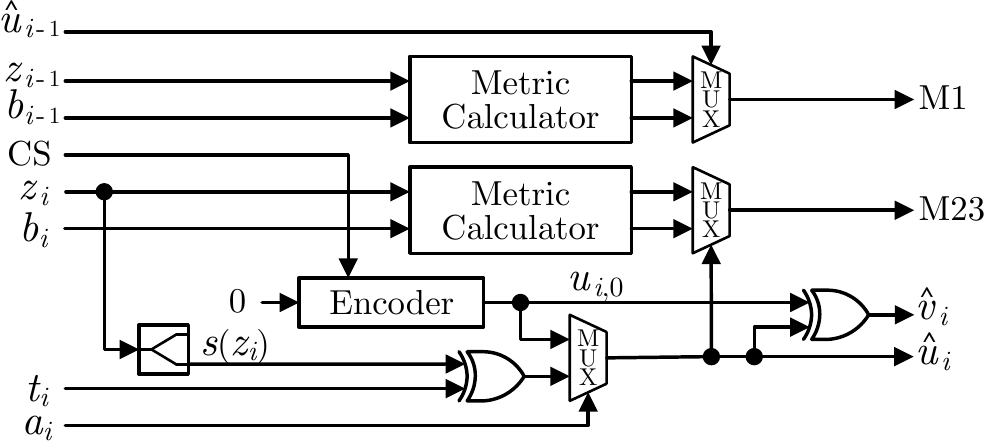}
\caption{Hardware diagram of branch metric unit (BMU).}
\label{fig_bmu} 
\end{figure}

Fig.~\ref{fig_bmu} shows the hardware diagram of BMU, which uses two metric calculator blocks to generate the current and previous branch metrics.
With a careful observation of Table \ref{tab_gamma} we realize that $\gamma_i(0) \geq \gamma_i(1)$ when $s(z_i) = 0$ and $\gamma_i(0) \leq \gamma_i(1)$ when $s(z_i) = 1$.
Hence, the most likely branch can be distinguished from the least likely branch without using an actual comparator.
The input $t_i$ is provided by FCU and is used to request the most likely branch metric (M2) when $t_i = 0$ or the least likely branch metric (M3) when $t_i = 1$ from BMU.
Additionally, when the current node N1 is frozen (i.e. $a_i = 0$) the BMU is forced to output the branch metric which corresponds to $\hat{v}_i = 0$.
We use a convolutional encoder to generate $u_{i,0}$ which is the convolution output for the assumption $\hat{v}_i = 0$.
In addition to M1 and M23 metrics, the BMU block provides FCU with the selected branch $\hat{v}_i$ and its corresponding convolution output $\hat{u}_i$.

\section{Implementation Results}\label{sec_results}
In this section, FPGA and ASIC implementation results of the proposed PAC Fano decoder are presented for block length $N = 128$ and message length $K = 64$.
We use $Q = 7$, $\Delta = 2$, $\mathbf{c} = (1,0,1,1,0,1,1)$, and choose $\mathcal{A}$ according to the Reed-Muller scoring rule as explained in \cite{arikan2019sequential}.
We use the hard quantized (1-bit quantization) values of bit-channel capacities \cite{moradi2020metric} as the bias vector $\mathbf{b}$.
The channel output LLR values are calculated at $E_b/N_0 = 3.5$ dB.

\subsection{FPGA Implementation Results}
The proposed PAC Fano decoder is successfully implemented onto Xilinx Nexys 4 Artix\textsuperscript{\textregistered}-7 ($28$ nm) FPGA.
The place-and-route results show that the decoder uses $16443$ lookup tables (LUTs) and $8306$ registers.
To evaluate the FER performance and measure the search complexity of the PAC Fano decoder, using MATLAB\textsuperscript{\textregistered} software, pseudorandom messages are generated, encoded, modulated using a binary phase-shift keying (BPSK) modulator, and transmitted to FPGA after white Gaussian noise is added.
The decoded carrier word $\hat{\mathbf{v}}$ is received from FPGA, $\hat{\mathbf{d}}$ is extracted from $\hat{\mathbf{v}}$ using $\hat{\mathbf{d}} = \hat{\mathbf{v}}_\mathcal{A}$, and compared with the actual transmitted message $\mathbf{d}$.
Also transmitted by the FPGA is the number of clock cycles consumed to decode each codeword which is measured by the CC counter.

\begin{figure}[t]
\centering
\includegraphics[width=\columnwidth]{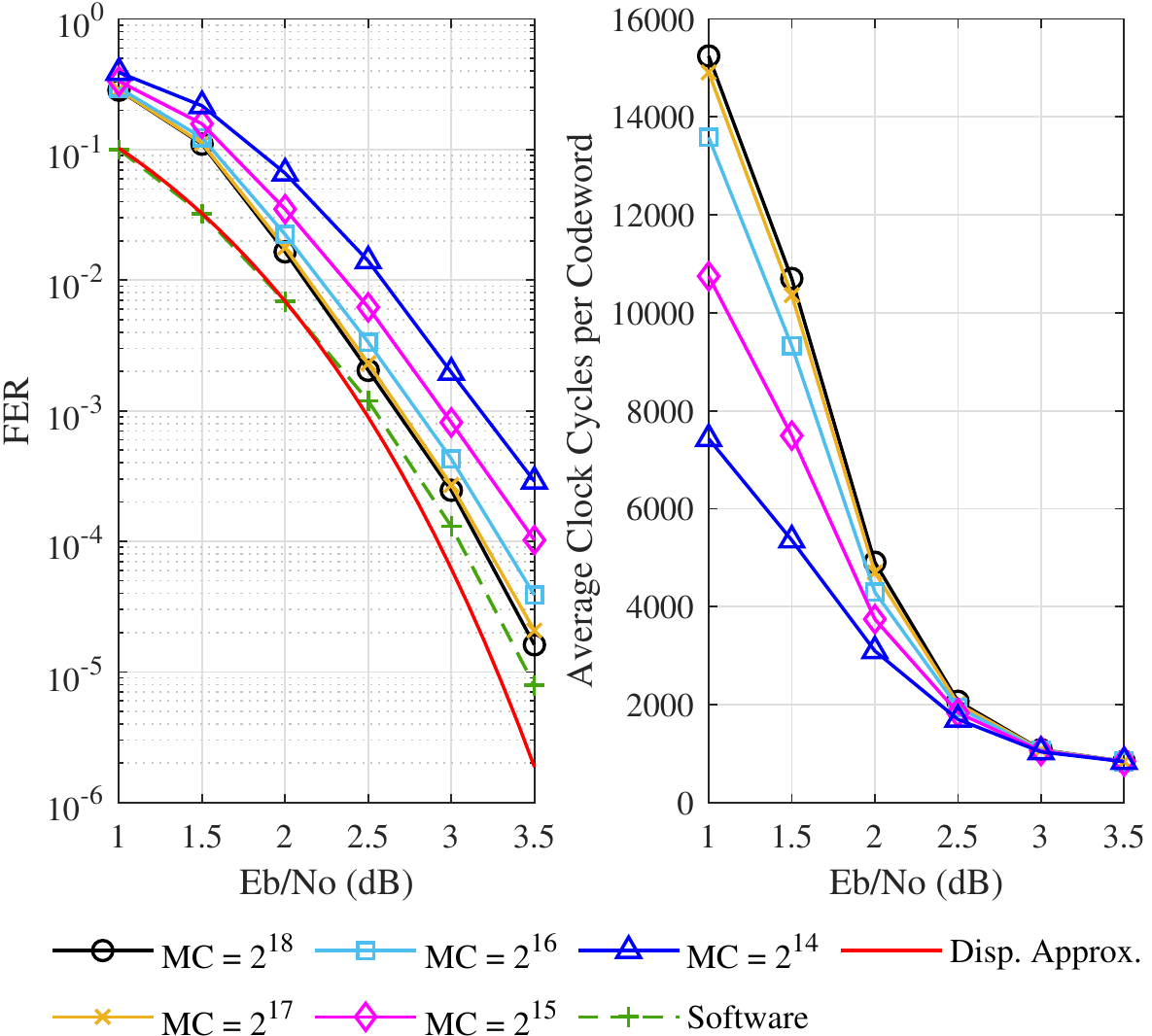}
\caption{Complexity and FER performance of PAC Fano decoder.}
\label{fig_FER_avgc}
\end{figure}

Fig.~\ref{fig_FER_avgc}~(left) plots the FER performance of the proposed PAC Fano decoder for different MC values.
The FER performance of software simulation of original PAC codes reported in \cite{arikan2019sequential} and the dispersion approximation are also plotted in this figure.
As expected, increasing the value of MC allows the Fano algorithm to perform more searches and maintain better FER performance.
With $\text{MC} = 2^{18}$, the proposed PAC Fano decoder obtains a FER performance close to the FER performance of software implementation at high SNR regime;
At $E_b/N_0 = 3.5$ dB the decoder achieves $\text{FER} = 1.6\times 10^{-5}$.
The performance loss is majorly due to the quantization of LLRs and approximation of the Fano metric.
In low SNR regime, the metric approximation error is large since the term $\log_2(1+e^{-(1-2\hat{u}_i) z_i})$ diverges from $|z_i|$ as SNR decreases.
But as SNR increases, this error becomes negligible.

Fig.~\ref{fig_FER_avgc}~(right) shows the average number of clock cycles consumed by the decoder for decoding a single codeword for different MC values.
The effect of MC value on average CC is significant at low SNR regime; as SNR increases, this effect fades out.
This is due to the Pareto distribution of Fano decoder's search complexity such that for high SNR values, only a small fraction of codewords require a very large search complexity \cite{moradi2020performance}.
With $\text{MC} = 2^{14}$, the average number of CCs per codeword drops by $51\%$ at $E_b/N_0 = 1$ dB at a cost of FER performance drop at high SNR values (significantly at $E_b/N_0 = 3.5$ dB).
At $E_b/N_0 = 3.5$, regardless of MC value, the decoder consumes an average of approximately $840$ CCs to decode a single codeword.
It is worth mentioning that for a noise-free channel (when no backtracking is done), the proposed PAC Fano decoder consumes $638$ CCs which corresponds to $5N-2$ CCs, of which $2N-2$, $2N$, and $N$ is consumed by PD, FCU, and BMU, respectively.

\subsection{Post-Synthesis Results}
Table~\ref{tab_asic_result} lists the post-synthesis results of the proposed PAC Fano decoder using Cadence\textsuperscript{\textregistered} Innovus\textsuperscript{TM} Implementation System with TSMC 28 nm 0.72 V library.
We present the results for the PAC Fano decoder with $\text{MC} = 2^{18}$.
The decoder occupies an area of 0.059 mm$^2$ and can operate at 500 MHz consuming 3.85 mW power.
The power value is estimated with Cadence\textsuperscript{\textregistered} Voltus\textsuperscript{TM} IC Power Integrity Solution using $10^4$ pseudorandom input vectors.
The performance values are reported at $E_b/N_0 = 3.5$ dB, and the average values are calculated from $10^7$ decoding trials.
The average information throughput (TP) of the decoder is estimated using $\text{TP} = (f_{\text{clk}}/\text{ACC})\times K$, where $f_{\text{clk}}$ is the operating frequency and ACC is the average number of CCs consumed for decoding a frame and is obtained from Fig.~\ref{fig_FER_avgc}.
The worst-case (W.-C.) latency of the decoder is determined by the value of MC.
The proposed PAC Fano decoder reaches an average information throughput of 38.1 Mb/s with an average latency of 839 CCs (1.68 $\mu$s).

\begin{table}[hbtp]
\centering
\caption{ASIC Implementation Results}
\label{tab_asic_result}
\begin{tabular}{lcc}
\hline
Technology                              & 28 nm\\
N                                       & 128 \\
K                                       & 64 \\ 
Supply Voltage (V)                      & 0.72 \\
\hline
Frequency (MHz)                         & 500 \\
Area (mm$^2$)                           & 0.059 \\
Power (mW)                              & 3.85 \\
\hline
Avg. Info. TP$^\dag$ (Mb/s)             & 38.1  \\
W.-C. Info. TP$^*$ (Mb/s)               & 0.12  \\
Avg. Latency$^\dag$ ($\mu$s)            & 1.68 \\
Avg. Latency$^\dag$ (CCs)               & 839 \\
W.-C. Latency ($\mu$s)                  & 524 \\
W.-C. Latency (CCs)                     & $2^{18}$ \\
\hline
Area Efficiency$^\dag$ (Mb/s/mm$^2$)    & 646 \\
Power Density (W/mm$^2$)                & 0.065 \\
Energy Efficiency$^\dag$ (PJ/bit)       & 101   \\
\hline
\multicolumn{2}{l}{$^\dag$Average value at $E_b/N_0 = 3.5$ dB.}\\
\end{tabular}
\end{table}

\section{Conclusion}\label{sec_conclusion}
In this brief, we proposed a hardware architecture for Fano decoding of PAC codes.
We introduced a new variant of Fano algorithm suitable for decoding PAC codes.
We also introduced a novel branch metric unit specific to PAC codes that can be implemented using simple logic gates.
Post-synthesis results showed that the decoder could provide an average information throughput of approximately 38 Mb/s at 3.5 dB with a power consumption of 3.85 mW and an area of 0.059 mm$^2$ for a block length of 128 and a code rate of $1/2$.
Due to its backtracking feature, the PAC Fano decoder has lower throughput than the state-of-the-art polar decoders (such as decoders of \cite{kestel}), but it exhibits better FER performance.
Because of their excellent FER performance at short block lengths and low encoding complexity, one of the potential use cases of PAC codes could be the Internet of Things (IoT), for which reliable communication is of great interest and low throughput and high decoding latency is tolerable.

\newpage
\bibliographystyle{IEEEtran}

\end{document}